\documentclass[twocolumn,twoside]{article}
\usepackage{graphics}
\newcommand{\h}{\linebreak \hspace*{3ex}}

\begin{document}
\title{
HIGH-RESOLUTION SPECTROSCOPY OF LONG-PERIODIC ECLIPSING BINARY $\varepsilon$ AURIGAE}
\author{Golovin Alex$^{1,2}$, Kuznyetsova Yuliana$^{2,3}$, Andreev Maxim$^{2,3}$\\[2mm] %English only
\begin{tabular}{l}
$^1$ Kyiv National Taras Shevchneko Univeristy, Kyiv, Ukraine \\ {\em golovin.alex@gmail.com, astron@mao.kiev.ua}\\
$^2$ Main Astronomical Observatory of National Academy of Science of Ukraine, Kyiv, Ukraine\\
$^3$ Terskol Branch of Institute of Astronomy RAS, Russia\\[2mm]
\end{tabular}
% no tabular, if one affiliation only
 }
\date{}
\maketitle
%Extended abstract not larger than 1 page
%Poster presentations - recommended volume 2-4 pages
%Oral presentations - recommended volume 2-4 pages
%Review paper - recommended volume 6-8 pages, exceptionally larger
%Please avoid large blank space in Your articles, if possible
%
% If the paper is submitted to another journal, but You plan to
% publish an (extended) abstract, please indicate something like
% "The complete paper is to be published in ..."
% The total width of the page is 170 mm=6.69inches= 2000 pixels (300dpi)
% The width of one column is 83 mm=3.26in=980 pixels (300dpi)
% The maximum height is 240mm=9.44in=2834pixels (300dpi)
% Please make tables and figures either for one (\begin{figure})
% or two (\begin{figure*}) dimension.
% Extensive tables may be printed with a \small font

ABSTRACT. The results of spectroscopic observations of
long-periodic eclipsing binary $\varepsilon$ Aur are reported. The
observations were carried out during 2 nights in 2007 at 2-meter
telescope located at the peak Terskol, Northern Caucasus (Russia).
Here we present series of $\varepsilon$ Aur spectra together with
EW measurements of the most prominent absorption lines. 
\\[1mm]
% Key words: Ierarchical structure according to the AAA list,
% the transition is marked by ":"
% different branches are separated by ";"
% same-level items are separated by ","
%\\[1mm]

{\bf 1. Introduction}\\[1mm]

  $\varepsilon$ Aur is well-observed long-periodic eclipsing binary, but still one of the most puzzling star.
It is the eclipsing binary with longest known orbital period --
27.1 year. The main enigma is considered in the eclipsing object
(it is supposed that the eclipsing body is of gigantic
proportions, on the order of 2,000 solar radii). Its nature
discussed for a long time, but still no reasonable explanation was
given.

The eclipsing nature of $\varepsilon$ Aur was first mentioned by
Fritsch (1824), where he discussed first ever-observed minimum in
1821. Since that $\varepsilon$ Aur' eclipses were observed each
27.1 years (Ludendorff, 1904) (in 1848, 1875, 1902, 1929, 1956,
1983), the next is expected in 2010 (first contact -- Aug, 06,
2009; mid-eclipse -- Jul, 09, 2010).

Recently (Carroll et al., 1991) $\varepsilon$ Aur
\textbf{\emph{secondary}} was interpreted as a protoplanetary
system. So, spectroscopic monitoring before and during the eclipse
is of great interest.
\\[1mm]

{\bf 2. Observations}\\[1mm]

  Spectroscopic observations were done at the Terskol Observatory (Russia, Northern Caucasus) during two nights, particularly at March, 30-31 and March, 31- April, 1 in 2007.
2-meter Zeiss telescope and coude-echelle spectrometer was used.
The wavelength range covers from 3660 to 9500 \AA in 80 orders.
The reciprocal dispersion ranges from 0.038 to 0.09 \AA/pixel. The
spectral resolution was R=45000.
\\[1mm]

\begin{table*}%[h]
 \caption{List of spectra and equivalent
widths of components of $\varepsilon$ Aur $H\alpha$ line.}
\begin{tabular}{cccccc}%{lrcl}
\hline
No. &  Exp. time, min &  S/N  & EW of blue wing, \AA & EW of central absorption, \AA & EW of red wing, \AA \\
\hline \hline \\ & \multicolumn{2}{l}{30.03.-31.03.2007}& \\
\hline
1  & 6 & 170 & -0.145 & 0.647 & -0.204 \\
2  & 6 & 200 & -0.154 & 0.645 & -0.209 \\
3  & 6 & 220 & -0.155 & 0.642 & -0.212 \\
4  & 6 & 250 & -0.155 & 0.644 & -0.211 \\
5  & 6 & 250 & -0.151 & 0.645 & -0.207 \\
6  & 6 & 220 & -0.155 & 0.643 & -0.210 \\
7  & 6 & 250 & -0.158 & 0.640 & -0.213 \\
8  & 6 & 220 & -0.156 & 0.641 & -0.208 \\
9  & 6 & 200 & -0.157 & 0.64  & -0.214 \\
10 & 6 & 230 & -0.155 & 0.642 & -0.213 \\
11 & 6 & 220 & -0.155 & 0.644 & -0.211 \\
12 & 6 & 200 & -0.154 & 0.642 & -0.212 \\
13 & 6 & 200 & -0.156 & 0.642 & -0.211 \\
14 & 6 & 220 & -0.154 & 0.642 & -0.214 \\
15 & 6 & 200 & -0.151 & 0.644 & -0.209 \\
16 & 6 & 200 & -0.153 & 0.645 & -0.211 \\
17 & 6 & 220 & -0.148 & 0.646 & -0.210 \\
18 & 6 & 200 & -0.154 & 0.644 & -0.214 \\
19 & 6 & 200 & -0.152 & 0.645 & -0.211 \\
20 & 6 & 200 & -0.145 & 0.645 & -0.193 \\
21 & 6 & 200 & -0.147 & 0.645 & -0.194 \\

\hline \hline \\ & \multicolumn{2}{l}{31.03.-01.04.2007}& \\
\hline
1  & 5 & 200 & -0.157 & 0.648 & -0.212 \\
2  & 5 & 200 & -0.156 & 0.645 & -0.208 \\
3  & 5 & 200 & -0.155 & 0.649 & -0.206 \\
4  & 5 & 200 & -0.157 & 0.647 & -0.207 \\
5  & 5 & 200 & -0.154 & 0.645 & -0.207 \\
6  & 5 & 220 & -0.155 & 0.648 & -0.206 \\
7  & 5 & 220 & -0.155 & 0.647 & -0.208 \\
8  & 5 & 200 & -0.156 & 0.646 & -0.205 \\
9  & 5 & 200 & -0.155 & 0.645 & -0.210 \\
10 & 6 & 180 & -0.156 & 0.646 & -0.208 \\
11 & 6 & 170 & -0.155 & 0.648 & -0.206 \\
12 & 6 & 180 & -0.153 & 0.649 & -0.200 \\
13 & 6 & 200 & -0.160 & 0.643 & -0.214 \\
14 & 5 & 200 & -0.155 & 0.650 & -0.201 \\
15 & 6 & 200 & -0.152 & 0.653 & -0.205 \\
16 & 6 & 220 & -0.157 & 0.650 & -0.212 \\
17 & 7 & 220 & -0.152 & 0.654 & -0.206 \\
18 & 7 & 200 & -0.149 & 0.655 & -0.204 \\
\hline \hline
\end{tabular}
\end{table*}

{\bf 3. Discussion}\\[1mm]

Our research was focused on searching for the short-time
variations of $H\alpha$ line profile. $H\alpha$ line was detected
in absorption together with prominent blue and red emission wings
symmetrical one to another, which is quite exciting (see Fig. 1
for the plot of $H\alpha$ region of $\varepsilon$ Aur spectrum).

%FIG.1****************************
\begin{figure}[h]
%\resizebox{8.26cm}{!}

\resizebox{\hsize}{!} {\includegraphics{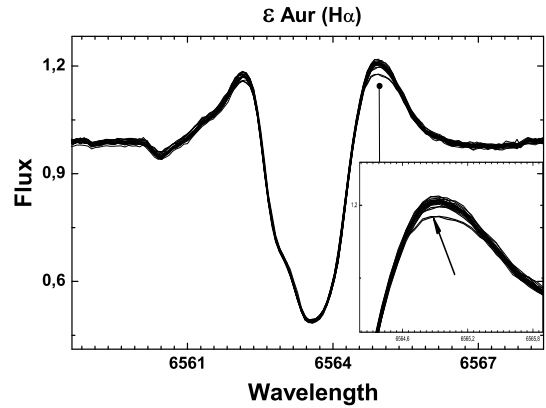}} \label{hh}
\caption{$H\alpha$ region of $\varepsilon$ Aur spectrum during
March, 30-31, 2007 observations}\end{figure}

The value variations of equivalent widths (EW) of the blue wing,
the red wing and the absorption core of the $H\alpha$ line
profiles were calculated ($F/Fc>1$ -- emission, $F/Fc<1$ --
absorption) and given in Table\,1.

 EW was calculated by direct numerical integration over the area
under the line profile.

As could be seen from Fig. 1, the blue wing of $H\alpha$ line
underwent a changes during the course of observations at March,
30-31, 2007. This changes reach up to 8\%, that could be
considered to be significant. During the next night no changes
were detected, reaching 5\% limit (see fig. 2 and 3 for plot of EW
changes in \% during both nights).

\begin{figure}[t]
%\resizebox{8.26cm}{!}

\resizebox{\hsize}{!} {\includegraphics{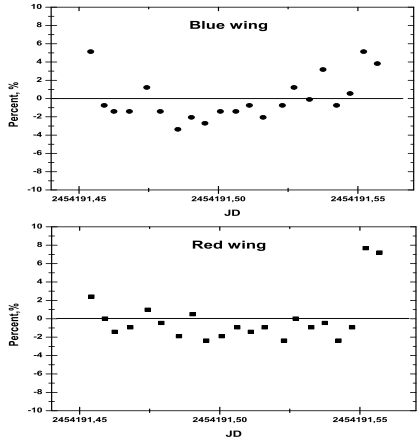}} \label{hh}
\caption{EW variability of $H\alpha$ during March, 30-31, 2007
observations}\end{figure}

\begin{figure}[h]
%\resizebox{8.26cm}{!}

\resizebox{\hsize}{!} {\includegraphics{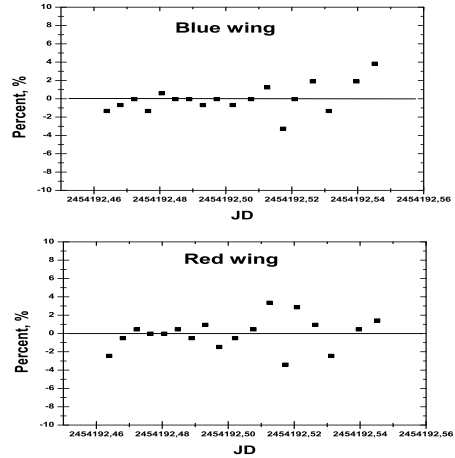}} \label{hh}
\caption{EW variability of $H\alpha$ during March, 31 -- April, 1 in
2007 observations}\end{figure}

\begin{figure}[h]
{\includegraphics{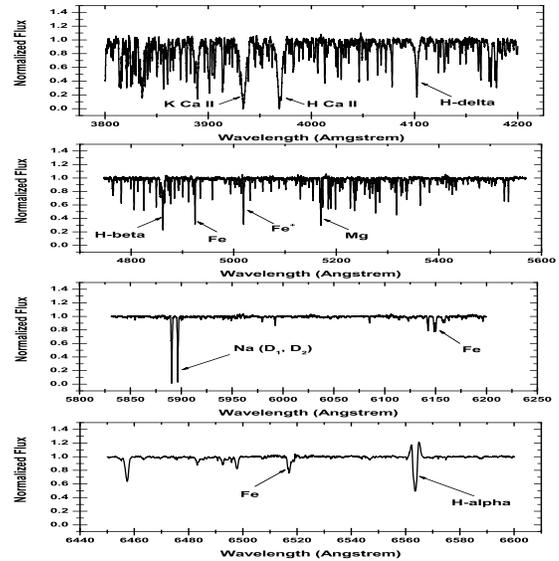}}% \label{hh}
 \caption{Portions of
$\varepsilon$ Aur average spectra, with the most prominent
absorption lines being marked}\end{figure}

Schanne, L. (2007) interpreted emission components of $H\alpha$ as
evidence of gas behind the star (for red-shifted component) and
radial outward flows, attributed to instabilities in the star
(blue-shifted component).

Cha et al. (1994), Cha et al. (1995) attributed blue wing emission
source to region region which contains an HII cloud with a short
time scale variation.

Also EW of the following absorption lines all of which exhibit
long-term variability (Thompson et al. 1987) were mesured: Fe\,I
(3922.9 \AA), Ti\,II (4028.0 \AA), Ti\,II (4443.85 \AA), Ti\,II
(4468.48 \AA), $H\beta$ (4861.5 \AA), Na\,DI (5889.953 \AA), Na\,DII
(5895.923 \AA), O\,I (7772\,\AA). No short-time  variability,
reaching 5\% limit, were detected.

Fig. 4 illustrates several portions of $\varepsilon$ Aur average
spectra, with the most prominent lines being identified and
denoted.

Further photometrical and spectroscopical monitoring of this
object is critically important for under- standing $\varepsilon$ Aur
 -- the most puzzling eclipsing binary.\\[1mm]

%\unitlength=1in
%\begin{figure}[h]
%\begin{picture}(3.26,3.14)% declared space X,Y for figure
%\vspace{3.14in}% this is instead of the figure file
% co-ordinates X,Y of the top left corner of the figure
% (to be centered)
%\put(X,Y){\special{em:graph figure1.pcx}}
%delete the remark sign "%" in the next line after changing
%the file name of Your figure and inserting Your X and Y
%\put(0,0.1){\special{em:graph figure1.pcx}}
%\end{picture}
%\caption{My model: $10^{80}$ white lines on the white background.}
%\end{center}
% the PostScript (EPS) figures may also be accepted. In this case, the
% beginning line should be
% \documentstyle[twocolumn,twoside,psfig]{article}
% instead of the mentioned above
%\end{figure}
%%%%% end of the figure

%\begin{figure}
%\resizebox{8.26cm}{!}
%\resizebox{\hsize}{!}
%{\includegraphics{Fig1.eps}}
%\label{hh}
%\caption{Light curve.}\end{figure}

%% begin of the table
% For the 2-column figure, use \begin{table*} \end{table*}
%% end of the table

\indent
{\bf References\\[2mm]}
Cha G. et al.: 1994, {\it A\&A}, {\bf 284}, 874-882. \\
Cha G. et al.: 1995, {\it IBVS}, No. 4149. \\
Fritsch J.M.: 1824, {\it Berl. Jahrb.}, p. 252. \\
Schanne L.: 2007, {\it IBVS}, No. 5747. \\
Thompson D.T. et al.: 1987, {\it ApJ}, {\bf 321}, 450-458.

% No number references, e.g. [1-18] are allowed. Author(s) and year, please
% the abbreviations of the journals may be according to the AAA standard,
% but longer (Astron. Zh. (AZh), Pis'ma Astron. Zh., Astrophys. J.)
% are also available
% If more than 3 authors, than may be shortened to Author1, 2, 3, et al.:
% electronic references are available, e.g.
% Schweitzer E.: 1999,\hc {\it ftp://cdsarc.u-strasbg.fr/pub/afoev/her/am}
% You may use either \h (enlarging the line from left to right)
% or \hc (no enlarging, no right align)
\vfill
%the final page should be formatted to make the height of the column
%nearly equal. This may be done by splitting the line in the first
%column e.g. by a command
%\linebreak\vfill\pagebreak\noindent
%
%Unfortunately, \thispagestyle{myheadings} command does not work in
%a new version of standard article.sty style file in the
%emtex\texinput directory. But You may restore the column title
% by simply replacing the line 193 in article.sty file
% \global\@topnum\z@ \@maketitle \fi\thispagestyle{plain}\@thanks
%by
% \global\@topnum\z@ \@maketitle \fi \@thanks

%Do not be afraid, we DID all this (and even more) formatting
%for previous volumes!
%Good luck !!!
\end{document}